\documentclass{article}

\PassOptionsToPackage{numbers,compress}{natbib}

\usepackage[nonatbib,final]{neurips_2020}

\usepackage[utf8]{inputenc} 
\usepackage[T1]{fontenc}    
\usepackage{hyperref}       
\usepackage{url}            
\usepackage{booktabs}       
\usepackage{amsfonts}       
\usepackage{nicefrac}       
\usepackage{microtype}      
\usepackage{amsmath}
\usepackage{xcolor}
\usepackage{natbib}
\usepackage{subfig}
\usepackage{graphicx}

\title{Towards Teachable Conversational Agents}

\author{%
  Nalin Chhibber\\
  Department of Computer Science\\
  University of Waterloo\\
  Waterloo, Canada \\
  \texttt{nalin.chhibber@uwaterloo.ca} \\
   \And
   Edith Law \\
   Department of Computer Science\\
   University of Waterloo\\
   Waterloo, Canada \\
  \texttt{edith.law@uwaterloo.ca} \\
}

\begin{document}

\maketitle

\begin{abstract}
The traditional process of building interactive machine learning systems can be viewed as a teacher-learner interaction scenario where the machine-learners are trained by one or more human-teachers.  
In this work, we explore the idea of using a conversational interface to investigate the interaction between human-teachers and interactive machine-learners.
Specifically, we examine whether teachable AI agents can reliably learn from human-teachers through conversational interactions, and how this learning compare with traditional supervised learning algorithms.
Results validate the concept of teachable conversational agents and highlight the factors relevant for the development of machine learning systems that intend to learn from conversational interactions.
\end{abstract}

\section{Introduction}

Recent progress in artificial intelligence has resulted in the development of intelligent agents that can direct their activities towards achieving a goal. Moreover, rapidly advancing infrastructure around conversational technologies has resulted in a wide range of applications around these agents, including intelligent personal assistants (like Alexa, Cortana , Siri, and Google Assistant), guides in public places (like Edgar \cite{fialho2013meet}, Ada and Grace \cite{traum2012ada}), smart-home controllers \cite{sciuto2018hey}, and virtual assistants in cars \cite{lugano2017virtual}.
This growing ecosystem of applications supporting conversational capabilities has the potential to affect all aspects of our lives, including healthcare, education, work, and leisure.
Consequently, agent-based interactions has attracted a lot of attention from various research communities \cite{cassell2000more, massaro1999developing, luger2016like, lopatovska2018talk, sciuto2018hey}. The success of these agents will depend on their ability to efficiently learn from non-expert humans in a natural way.

In this paper, we explore the idea of using conversational interactions to incorporate human feedback in machine learning systems.  We evaluate this concept through a crowdsourcing experiment where humans teach text classification to a conversational agent, with an assumption that the agent will assist them with annotations at a later time. Overall, this paper contributes towards a larger goal of using conversations as a possible interface between humans and machine learning systems with the following key contributions:
\begin{itemize}
    \item The idea of leveraging conversational interactions to tune the performance of machine learning systems, which can be extended to personalize assistants in future.
    \item An interactive machine learning algorithm that learns from human feedback, and considers statistical as well as user-defined likelihood of words for text classification.
\end{itemize}

\section{Related Work}
Traditional machine learning systems that only make data-driven predictions, tend to be as good as the quality of the training data. However, the data itself may suffer from various biases and may not accurately represent all human-specific use-cases. 
{\it Interactive machine learning} attempts to overcome this by involving users in the process of training and optimizing the machine learning models. By allowing rapid, focused and incremental updates to the model, it enables users to interactively examine the impact of their actions and adapt subsequent inputs.
In essence, interactive machine learning is a way to allow meaningful human feedback to guide  machine learning systems. One of the earliest work in this area is from Ankerst et al. who worked on an interactive visualization of classification tree \cite{ankerst1999visual}. They created an interface that provide sliders to adjust the number of features or threshold values for each node in the decision tree, and interactively display the classification error.
Ware et al. \cite{ware2001interactive} demonstrated that humans can produce better classifiers than traditional automatic techniques when assisted by a tool that provides visualizations about the operation of specific machine learning algorithms.
Fails and Olsen studied the difference between classical and interactive machine learning and introduced an interactive feature selection tool for image recognition \cite{fails2003interactive}. Von et al. introduced ReCAPTCHA as a human computation system for transcribing old books and newspapers for which OCR was not very effective \cite{von2008recaptcha}. Fiebrink et al. created a machine learning system that enable people to interactively create novel gesture-based instruments \cite{fiebrink2011human}. Their experiments found that as users trained their respective instruments, they also got better and even adjusted their goals to match observed capabilities of the machine learner. These examples illustrate how rapid, focused and incremental interaction cycles can facilitate end-user involvement in the machine-learning process. 
Porter et al. \cite{porter2013interactive} formally breaks down the interactive machine-learning process into three dimensions: task decomposition, training vocabulary, and training dialogue. These dimensions define the level of coordination, type of input, and level/frequency of interaction between the end-users and machine learners.  Later, Amershi et. al examined the role of humans in interactive machine learning, and highlighted various areas where humans have interactively helped machine learning systems to solve a problem \cite{amershi2014power}. Their case study covered various situations where humans were seen as peers, learners, or even teachers while engaging with interactive systems across different disciplines like image segmentation and gestured interactions. A form of interactive machine learning has been studied under apprenticeship learning (also learning by watching, imitation
learning, or learning from demonstration) where an expert directly demonstrate the task to machine learners rather than telling them the reward function \cite{abbeel2004apprenticeship}. However, this is tangential to our current work as we specifically focus on providing active guidance through conversational interactions instead of passive demonstrations. 

A special case of interactive machine learning is \textit{active learning} which focuses on improving machine learner's performance by actively querying a human oracle and obtain labels \cite{settles2009active}. However, several studies reveal that active learning can cause problems when applied to truly interactive settings \cite{cakmak2010optimality, cakmak2010designing, guillory2011simultaneous}. 
Simard et al.~\cite{simard2017machine} formalize the role of teachers as someone who transfer knowledge to learners in order to generate useful models.  Past work on algorithmic teaching shows that while human teachers can significantly improve the learning rate of a machine learning algorithm \cite{balbach2009recent, mathias1997model, goldman1995complexity}, they often do not spontaneously generate optimal teaching sequences as human teaching is mostly optimized for humans, and not machine learning systems. Cakmak et al. examined several ways to elicit good teaching from humans for machine learners \cite{cakmak2014eliciting}. They proposed the use of teaching guidance in the form of an algorithms or heuristics. Algorithmic guidance is mostly studied under algorithmic teaching \cite{balbach2009recent}, and aims to characterize teachability of concepts by exploring compact (polynomial-size) representation of instances in order to avoid enumerating all possible example sequences. On the other hand, heuristic-based guidance aims to capture the intuition of an optimal teacher and enable them to approximate the informativeness of examples for the learner. While algorithmic guidance can provide guaranteed optimality bounds, heuristic-based guidance is easier to understand and use \cite{cakmak2014eliciting}. Consequently, recent work in this area has started focusing on the human-centric part of these interactive systems, such as the efficacy of human-teachers, their interaction with data,
as well as ways to scale interactive machine learning systems with the complexity of the problem or the number of contributors \cite{simard2017machine}. 
However, these solutions have not been studied within the context of conversational systems.




\section{System Description} \label{section:system_description}
In this work, we introduce a teachable agent that learns to classify text using human feedback from conversational interactions. In this section, we describe the task environment used for teaching the agent, architecture of the dialog system and finally the learning mechanism of the agent.

\subsection{Task Environment}

\begin{figure}[h!]
    \centering
    \subfloat[Teaching Interface]{{\includegraphics[height=4.5cm]{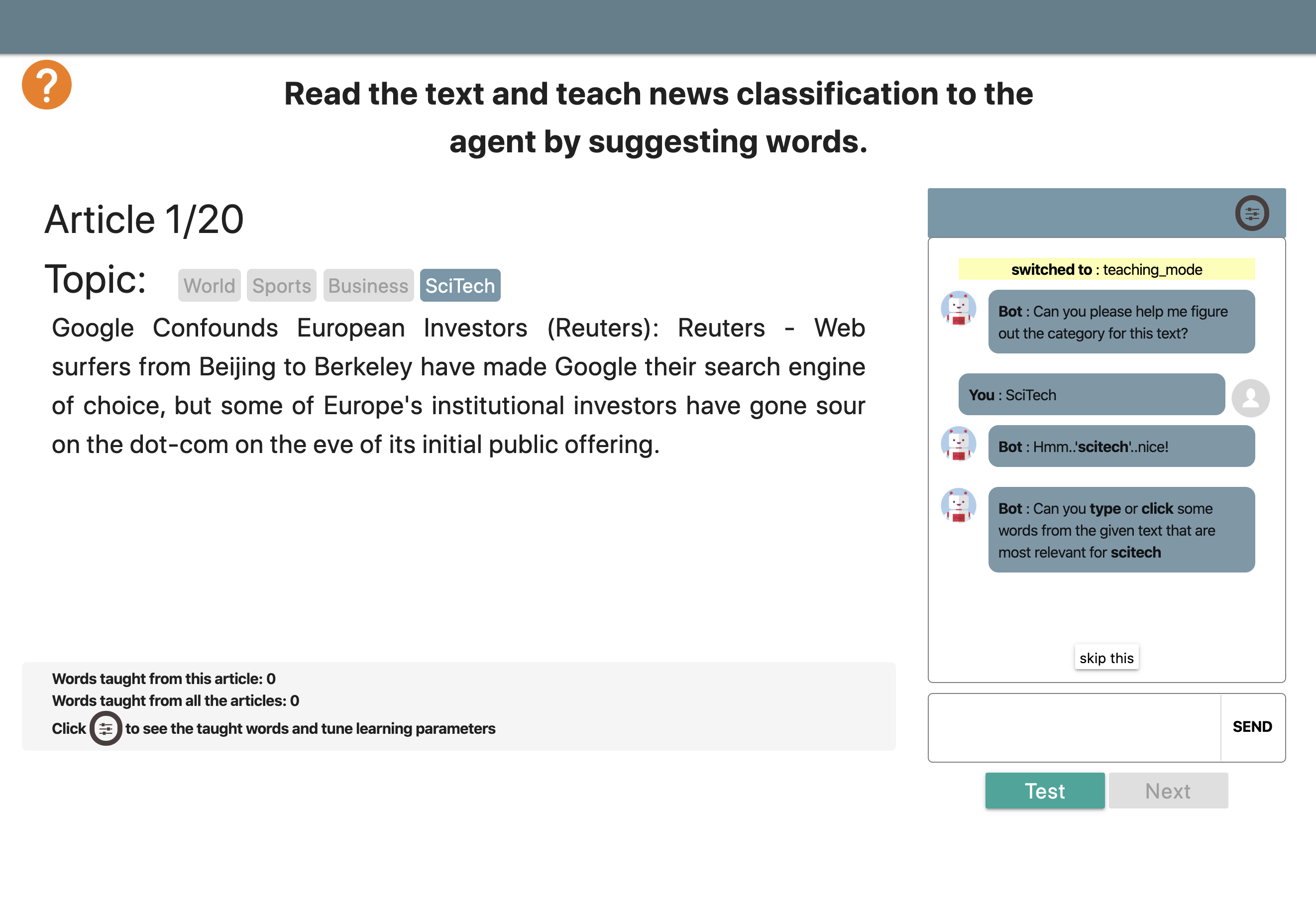} }}%
    \qquad
    \subfloat[Testing Interface]{{\includegraphics[height=4.5cm]{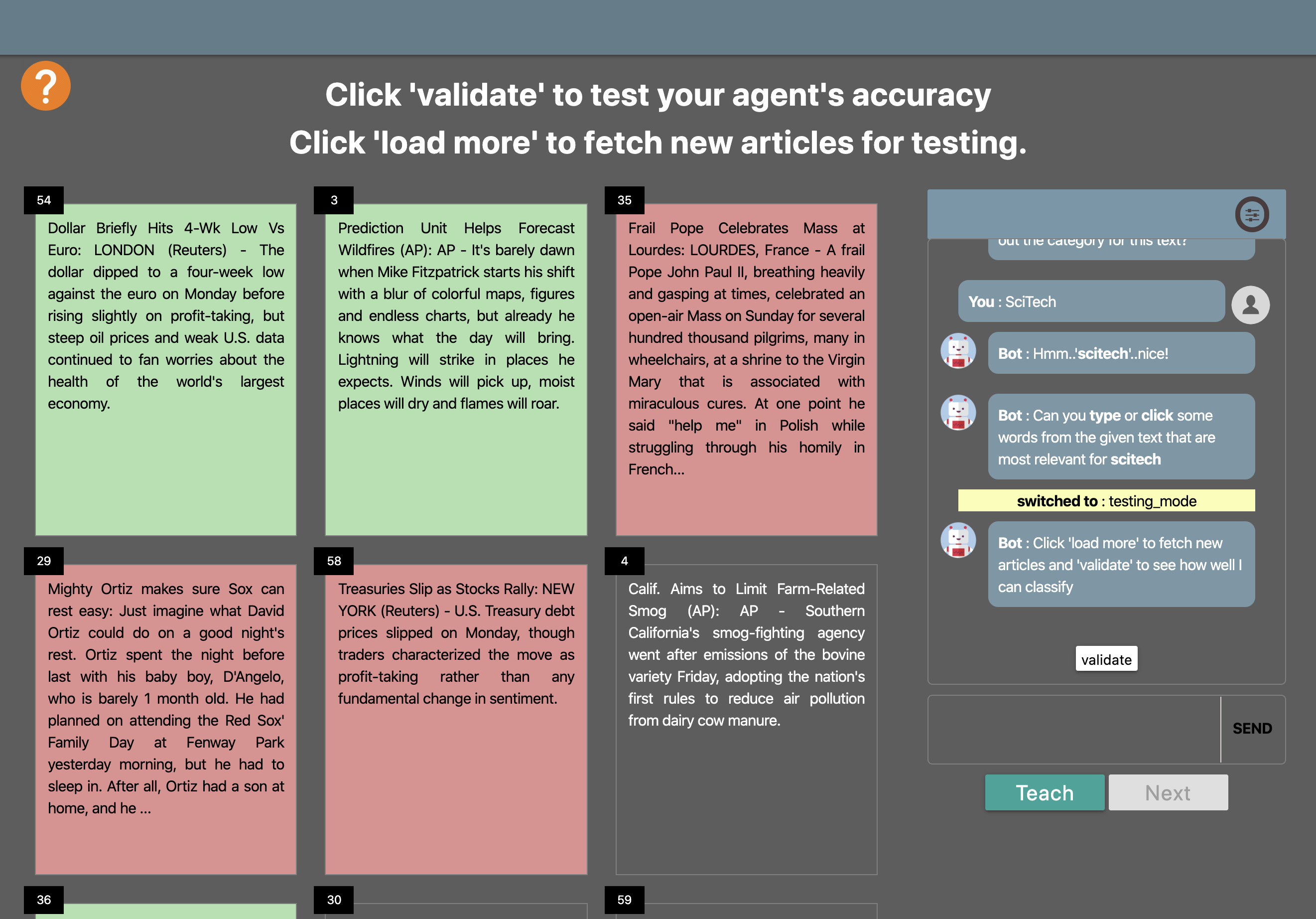} }}%
    \caption{Task Environment}%
    \label{fig:system_interfaces}%
\end{figure}



The teachable agent was deployed as a textual conversational bot embedded into a web-based learning environment. In the task interface, participants read an article and converse with the conversational agent to teach them how to classify that article. There were two modes, teaching and testing, as described in Figure \ref{fig:system_interfaces}. In the teaching mode, while reading the article, participants could enter or highlight words to explain why an article should be classified in a particular way (Figure \ref{fig:system_interfaces}a). The agent asked questions to the human-teacher and revealed what it did not understand about the topic, or what else it wanted to know. In answering the agent's questions, the human teachers were prompted to reflect on their own knowledge. The assumption was that through this process, human teachers may gain a better understanding about how to perform the classification task themselves. Every human teacher taught their own agent. In the testing mode, participants could present new articles to the teachable agent, and ask them to classify articles in real-time based on what they have learned from the conversational interaction (Figure \ref{fig:system_interfaces}b). After the agent's prediction, correctly classified articles were coloured green by the system, whereas incorrectly classified articles were coloured red. During the entire interaction, participants were encouraged to frequently test the agent to assess their teaching performance and how well the agent was handling unseen examples.

\subsection{Dialog System}
\label{section:conversational_interface}

\begin{figure}[h!]%
    \centering
    \subfloat[Teaching]{{\includegraphics[height=7cm]{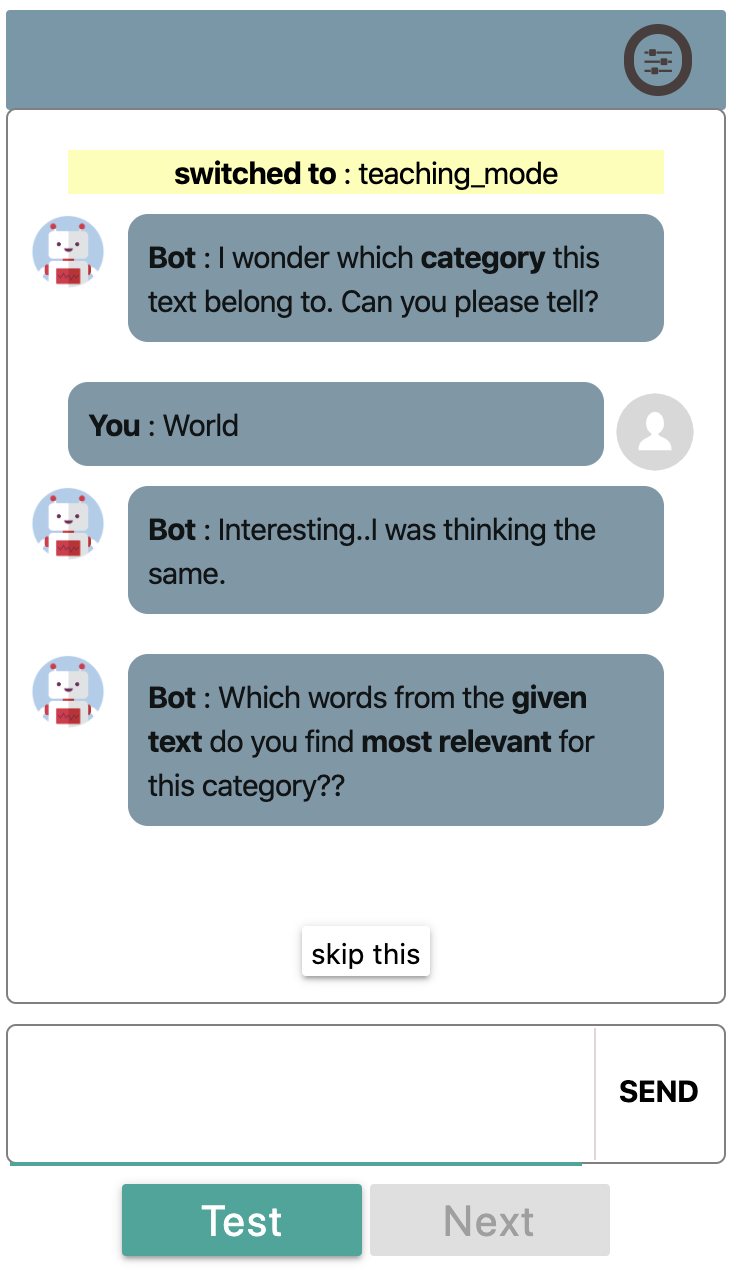} }}%
    \qquad
    \subfloat[Testing]{{\includegraphics[height=7cm]{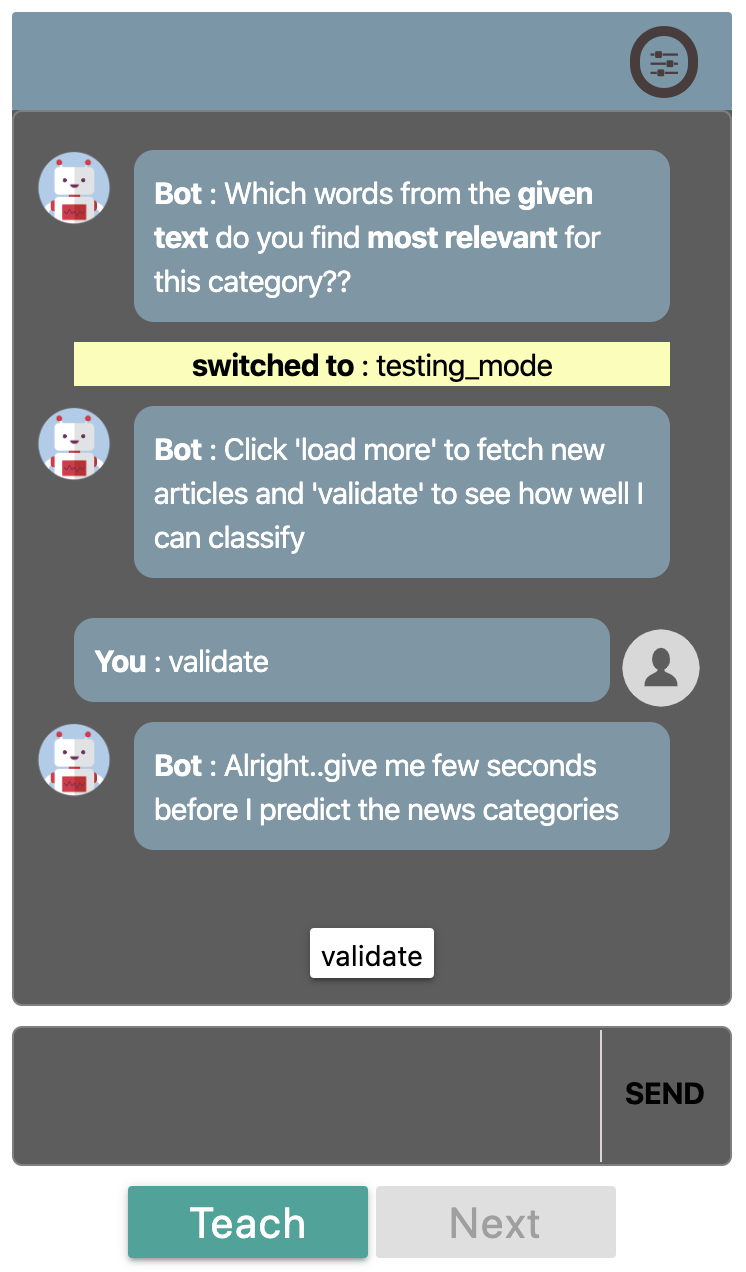} }}%
    \caption{Interaction with the agent during (a) teaching, and (b) testing mode}%
    \label{fig:exp_1_teach_test}%
\end{figure}

Agent's dialogue system was designed using conversational tree, a branching data structure where each node represents a place where a conversation may branch, based on what the user says ~\cite{adams2014fundamentals}. Edges in a conversational tree can be traversed backward or forward because of the nature of conversational interaction; for example, the traversal is backwards if the agent is asked to repeat a sentence.  Besides the conversational tree, the state of the conversation was maintained using a hierarchical state machine. The top-most level of this hierarchy was the split between the learning and testing modes.  In the learning mode, the teachable agent was focused on learning new features through conversations related to a given topic; whereas in the testing mode, agent predicted the category of unseen articles and asked for more samples from the human teachers. Each of these modes further contained multiple contexts that defined the agent's current understanding about the relevance of features. The agent could switch between different contexts in order to capture new features that were relevant or irrelevant to the topic under discussion. This switching between different contexts was made possible by explicit user actions, as well as intent identification.
For the latter, we used a rule-based approach to identify different intents during the conversational interactions. In addition, we also developed agent strategies loosely consistent with Speech Act theory \cite{searle1980speech}, that directs the user to ask about content within the agent's dialog system repertoire. In certain cases in which no input was recognized, the agent would default to one of several fallback options like: asking users to paraphrase, repeat or simply ignore and move to next article. Sample conversational interaction during the teaching and testing modes are shown in Figure \ref{fig:exp_1_teach_test}A and \ref{fig:exp_1_teach_test}B respectively.

\begin{table}
  \caption{Three types of heuristic teaching guidance}
  \label{table:heuristic}
  \centering
  \begin{tabular}{p{2.8cm}|p{4.3cm}|p{5.7cm}}
    \toprule
    \cmidrule(r){1-2}
    \textbf{Heuristic}  & \textbf{Description} & \textbf{Conversational Guidance} \\
    \midrule
    Externally relevant words & Words 'outside' the text that will most likely describe the $category$ & Can you tell me few more words that should describe the $category$ but are not in the text?\\
    Internally relevant words & Words from the text that are most relevant to the $category$ & I wonder which words are most relevant while categorizing this text to the $category$?\\
    Internally irrelevant words & Words from the text that are least relevant to the $category$ & Which words are least relevant while categorizing this text to the $category$?  \\
    \bottomrule
  \end{tabular}
\end{table}

Table \ref{table:heuristic} summarizes the different types of heuristic teaching guidance that the human teacher can provide.  We identified these three teaching heuristics based on Macgregor et al.~\cite{macgregor1988effects}, who proposed teaching heuristics for optimizing the classification algorithms.  Features identified through these heuristics were meant to supplement the classifier by proposing new features, amplifying relevant ones, or discounting the irrelevant ones.

\subsection{Agent's Learning Mechanisms}

The agent learns to classify articles using an enhanced version of the Naive Bayes algorithm that incorporates human teaching as additional input.  Naive Bayes is a generative classifier, which computes the posterior probability $P(y|x)$ (i.e., the probability of a class $y$ given data $x$); for text classification, the assumption is that the data is a bag of words and that presence of a particular word in a class is independent to the presence of other words in that class. One advantage of Naive Bayes, especially in the context of interactive teaching, is that it can be trained quickly and incrementally.
Formally, the Naive Bayes model can be expressed as:
\begin{equation}
\begin{split}
P(C_k|w_1, w_2...w_n) 
                       &  \propto P(C_k) \prod_{i=1}^{n} P(w_i|C_k)
\end{split}
\label{eq_nb_proportional_}
\end{equation}

Here the variable $C_k$ represents a document class from (World, Sports, Business, or SciTech) and $\textbf{w} = (w_1, w_2, w_3...w_n)$ are the individual words from the respective document.
Naive Bayes is known to perform well for many classification tasks even when the conditional independence assumption on which they are based is violated \cite{domingos1996beyond}. However, many researchers have tried to boost their classification accuracy by relaxing this conditional independence assumption through locally weighted learning methods \cite{atkeson1997locally,  frank2002locally}.
We adopt a similar idea of relaxing the feature independence assumption by considering the relevant and irrelevant features ({\it conversational keywords}) that a human teacher mentions during interaction on a particular topic. We infer the class of a test document by considering its constituent words, as well as {\it similar} conversational keywords captured from the teaching conversation. Given the set of words in a test document, the conditional probability for those words in training data under respective classes is represented as $P(w_i|C_k)$ and the conditional probability of conversational keywords that are similar to the words in the corpus is represented as $P(s_i|C_k)$.
\begin{equation}
\begin{split}
 P(s_{i}|C_k) & = \frac{\mbox{\# conversational keywords similar to $word_{i}$ in test document for $C_k$}}{\mbox{Total \# conversational keywords captured from the interaction for $C_k$}}
\end{split}
\label{eq_nb_similarity}
\end{equation}
To determine the similarity between conversational keywords and words from the test document, we used the cosine similarity of their vector representations as a proxy for semantic closeness. Cosine similarity has a range between -1 and 1, with negative values indicating dissimilar word vectors, and positive values indicating greater similarity between the word vectors. These word vectors were obtained using Word2vec model: a shallow neural-network that is trained to reconstruct the linguistic contexts of words in vector space \cite{mikolov2013distributed}. We used 300-dimension word vectors trained on 300,000 words from Google News dataset, cross-referenced with English dictionaries. Conversational keywords where the similarity coefficient is below a threshold (e.g. 0.2) were not considered in \eqref{eq_nb_similarity}. Having determined the set of conversational keywords that are similar to the document words, we modify the posterior probability in two different ways:

{\it Case 1: Without supervised pre-training}.  
In this case, the posterior probability is inferred only from the conditional probability of the conversational keywords captured during teaching. Thus, equation \eqref{eq_nb_proportional_} can be expressed as:
\begin{equation}
\begin{split}
P(C_k|w_1, w_2...w_n, s_1, s_2...s_n) & \propto P(C_k) \prod_{i=1}^{n} P(s_i|C_k)
\end{split}
\label{eq_nb_userdefined}
\end{equation}

{\it Case 2: With supervised pre-training}.  
In this case, the posterior probability is inferred from both the conditional probability of the conversational keywords captured during teaching and the conditional probability of the words in the original corpus. Thus, equation \eqref{eq_nb_proportional_} can be expressed as:
\begin{equation}
\begin{split}
P(C_k|w_1, w_2...w_n, s_1, s_2...s_n) & \propto P(C_k) \prod_{i=1}^{n} P(w_i|C_k) P(s_i|C_k)
\end{split}
\label{eq_nb_combined}
\end{equation}
Note that the conditional probability of a word appearing in the training corpus, $P(w_i|C_k)$, and the conditional probability of similar words being discussed during the conversational interaction, $P(s_{i}|C_k)$ are considered as two independent events and hence their combined probabilities can be expressed as the product of individual probabilities. 
To get the final classification, we output the class with highest posterior probability.  For equation \ref{eq_nb_userdefined} and equation \ref{eq_nb_combined}, this can be calculated as:
    $y = \underset{}{\mathrm{argmax}}\   P(C_k) \prod_{i=1}^{n} P(s_i|C_k)$
, and
$y = \underset{}{\mathrm{argmax}}\   P(C_k) \prod_{i=1}^{n} P(w_i|C_k)\ P(s_i|C_k)$ respectively

\section{Experiment} \label{section:formative_evaluation}
We conducted a formative experiment to investigate whether humans can interactively teach text classification to conversational agents. We validate this by comparing the performance of the underlying Naive Bayes algorithm with and without using the supervised pre-training, as well as against baseline text classification algorithms with no human feedback.

\subsection{Procedure}
We recruited sixty crowdworkers from Amazon Mechanical Turk (10 females, 50 males), 23 to 53 years old (M= 30.9, SD= 5.29). The study was conducted by posting Human-Intelligence-Tasks (HITs) with the title: ``Teach How to Classify News Articles to a Chatbot''. 
87\% of the participants were native English speakers, but all reported some prior experience with conversational agents on a 7-point scale (M=5.76, SD=1.15). 53.4 \% of the participants reported prior experience in teaching a classification task to someone else, the other half had no prior experience in teaching (46.6\%).  Regarding the prior knowledge on the given news categories, participants rated most for World (M=5.85, SD=1.20), followed by SciTech (M=5.63, SD=1.27), Business (M=5.55, SD=1.47) and Sports (M=5.07, SD=1.78).
The experiment took approximately 20-30 minutes to complete.



After accepting the HIT, providing consent, and completing the demographic questionnaire, participants were given a short tutorial on the task interface. 
During the main phase of the experiment, there were 20 articles to teach that were equally distributed across all four news categories. Participants were supposed to teach at least one word from each article and were also allowed to switch between teaching and testing modes in order to check their agent's performance. 
During the teaching process, the agent asked questions that participants would answer in order to teach them how to classify articles into one of the four categories. In the test mode, the agent would predict the category of unseen articles based on words that were taught during the teaching interaction. Participants were free to switch between the "Teach" and "Test" modes by clicking respective buttons below the chatbox.


Articles for text classification were sampled from a subset of AG News Classification Dataset~\cite{zhang2015character}, with 4 largest classes representing the topics World, Sports, Business and SciTech. Each class contained 30,000 training samples and 1,900 testing samples. The total number of training samples in the dataset were 120,000 and number of test samples were 7,600.  We used the standard data pre-processing techniques including tokenization, stop-words removal and lemmatization. Tokenization was done using $word\_tokenize()$ function from NLTK that splits the raw sentences into separate word tokens. This was followed by a text normalization step where we converted individual tokens into lowercase to maintain the consistency during training and prediction. Stop-words filtering was also done using NLTK to filter out the words that did not contain vital information for text classification. Finally, we used WordNetLemmatizer with part-of-speech tags to obtain the canonical form (lemmas) of the tokens. Conversion of tokens to their base form was done to reduce the language inflections from words expressing different grammatical categories including tense, case, voice, aspect, person, number, gender, and mood.

\subsection{Results}
Throughout the study, a total of 31,199 dialogues were exchanged between sixty users (12,020) and the conversational agent (19,179), with an average of 520 total dialogues per session (200.3 by the user and 319.6 by the conversational agent). 
Average F1-score of the agent was recorded as 0.48 (SD= 0.15). 
Participants' background did not show any significant impact on their agent's F1-score, but as the number of dialogues exchanged by the participants increased, their agent's performance also significantly increased, $\beta= 0.001, t(56)= 3.68, p < 0.001$. 
Native English speakers tend to speak more than non-native English speakers throughout the experiment, $\beta= -0.21, t(54)= -2.04, p =.05$.
There was also a significant increase in the F1-score with increasing number of agent testing, $\beta= 0.005, t(56)= 4.69, p < 0.001$. However, the overall F1-score seemed to decrease when more external words was taught, $\beta= -0.0003, t(55)= -2.16, p = 0.03$.

\begin{figure}[h!]
\begin{center}
  \includegraphics[width=\linewidth]{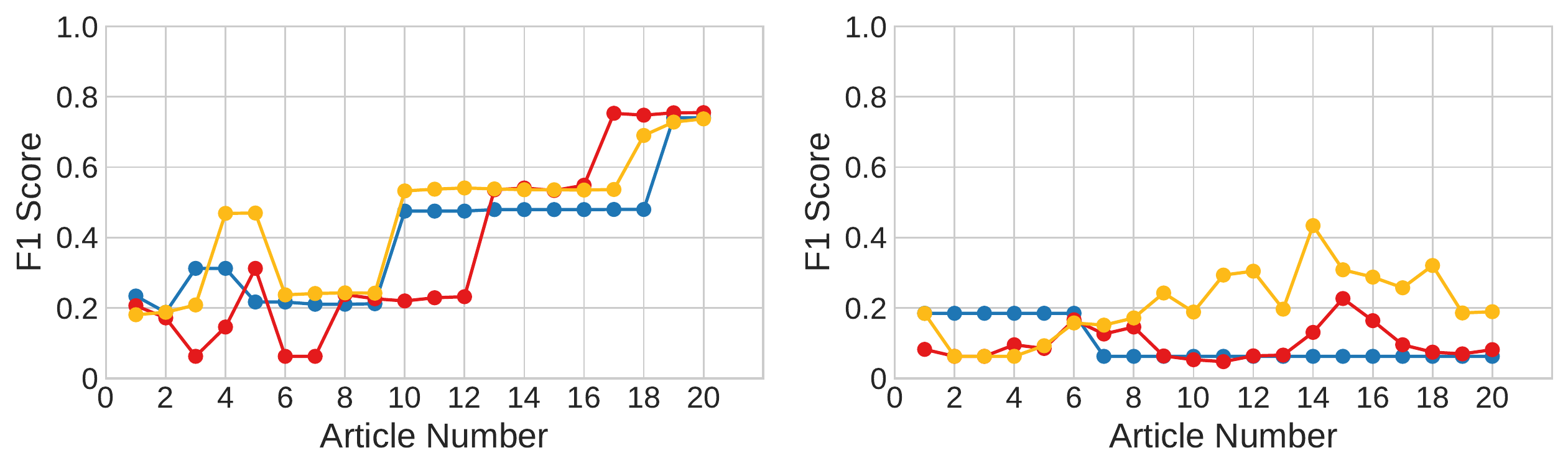}
  \end{center}
  \caption{Change in F1-scores of the agent when taught by 3 (a) most successful, (b) least successful crowdworkers, with no supervised pre-training of the interactive Naive Bayes classifier}
  \label{fig:exp_1_good_bad_teacher}
\end{figure} 

We calculated the classification performance of the agent after each news article that was discussed during the conversational interaction. Although the classifier was trained online on the keywords captured from conversations on an article, along with the keywords captured from all previous conversations, the performance was calculated ``offline'' on the entire test set of 7600 articles from the AG News Dataset treating individual article as an epoch. For this, we used the interactive variant of Multinomial Naive Bayes classifier as described in equation \eqref{eq_nb_userdefined}. Since the classifier was used without supervised pre-training, the initial performance was around 20\% before the interaction. After the interaction, some of the most successful crowdworkers were able to increase the performance of the agent to around 70\%, while for the least successful ones, the performance decreased to ~10\%. Results indicate that the final performance of classifier varied significantly across different participants. We did not find a direct co-relation between the number of words taught and the classification performance. This indicates that the quantity of the words captured alone does not impact the classifier's performance. Figure \ref{fig:exp_1_good_bad_teacher} shows the progression of F1-score with each article for three most successful and least successful teachers, that trained an interactive machine learner without supervised pre-training. 

\begin{table}[ht]
\caption{Comparison of baseline classifiers with interactive variants of Naive Bayes with supervised pre-training, for best teacher, worst teacher and all teachers. }
\label{table:benchmark}

\begin{tabular}{ p{6cm}|p{2cm}|p{2cm}|p{2cm}  }
 \toprule
 \textbf{Model}  & \textbf{Precision} & \textbf{Recall} & \textbf{F1-Score} \\
 \hline
 \multicolumn{4}{c}{Without Teachers (Baseline)}\\
 \hline
 Bernoulli Naive Bayes & 0.8626  & 0.8584  & 0.8593 \\
 Multinomial Naive Bayes & 0.8899 & 0.8902  & 0.8900 \\
 \hline
 \multicolumn{4}{c}{Best Teacher}\\
 \hline
 Interactive Bernoulli Naive Bayes  & \textbf{0.8658}   &  \textbf{0.8672} &  \textbf{0.8664} \\
 Interactive Multinomial Naive Bayes  & \textbf{0.8972}  &  \textbf{0.9042} &  \textbf{0.9006}\\
 \hline
 \multicolumn{4}{c}{Worst Teacher}\\
 \hline
 Interactive Bernoulli Naive Bayes   & 0.8145   &  0.8247   & 0.8196  \\
 Interactive Multinomial Naive Bayes   &  0.8729  &  0.8709  &  0.8719\\
 \hline
 \multicolumn{4}{c}{All Teachers}\\
 \hline
 Interactive Bernoulli Naive Bayes  & 0.8532   &  0.8578   & 0.8558  \\
 Interactive Multinomial Naive Bayes  & 0.8847   &  0.8830  & 0.8838  \\
 \bottomrule
\end{tabular}
\end{table}



Next, we investigated the results of classifier's performance for other interactive variants of Naive Bayes with supervised pre-training as described in equation \eqref{eq_nb_combined}. These results were obtained "offline", by simulating the learning conditions after the experiment. Both statistical likelihood of words from relevant classes, and the user-defined likelihood obtained from conversations were used to calculate the posterior probability of test-documents. The classification performance of the interactive variants of Naive Bayes were compared with the two baselines for Bernoulli Naive Bayes and Multinomial Naive Bayes respectively. The comparison was made between most successful, least successful, and combination of all crowdworkers who taught the teachable agent during the experiment. Surprisingly, combined effect of teaching from all the participants seemed to reduce the overall performance of the learner in an interactive conversational setting. Precision, recall and F1 scores for all interactive variants are described in Table \ref{table:benchmark}.

\section{Discussion}

In this work, we described the concept of leveraging conversational interactions as an interface between humans and an interactive machine learning system.
It was found that performance of the agent improved with increase in the number of dialogues exchanged by participants and the number of times it was tested during the session. This implies that participants who were concerned about their agent's performance through repeated testing were more successful in training the agent on news classification task. Further, classification performance of the agent seem to degrade when they were taught more external words that were outside the given article. 
An interesting finding is that the combined effect of teaching from all the crowdworkers may actually reduce the overall performance of the learner in an interactive conversational setting (Table \ref{table:benchmark}). This indicates that learning from a lot of sources may affect the performance of the learner if the proportion of ineffective teachers is significantly more than effective ones, and teaching from effective and ineffective sources is not easily distinguishable. 
It was also observed that native English speakers tend to exchange more dialogues throughout the experiment. This implies that localization of dialogue systems is useful for longer engagement.

\subsection{Limitations and Future Work}
The performance of our proposed interactive machine learning algorithm is based on the cosine similarities obtained from the vector representation of words. We used a compressed variant of Word2Vec trained on a smaller dataset due to performance reasons, which limits the quality of word embeddings used. Future investigations can focus on contextual embeddings (like BERT) trained on more relevant and richer dataset for better outcomes. Further, results from the experiment shows that effective human teaching leads to better machine-learners. However, it remains unclear what characteristics are specific to a good teacher and which factors influence the quality of teaching. 
Moreover, it will be interesting to explore different modalities of the interaction with teachable agents as opposed to a textual conversational interaction. Follow up experiments may involve the use of voice-based agents or embodied agents like physical robots to validate the results in different contexts. 
Finally, while the proposed algorithm focuses on transparency by using Naive Bayes classifier as the baseline machine learning model, it remains unclear how the idea of teachable conversational agents will extend to state-of-the-art systems. Future work can investigate how human feedback through conversational interactions can be used to improve machine learners based on modern deep learning architectures.

In conclusion, this paper aims to take one step in the direction of building teachable conversational agents and how they learn from human teachers. Understanding various nuances across these facets will be useful for building interactive machine learners that aim to reliably learn through conversational interactions.

\bibliographystyle{unsrtnat}
\bibliography{_references}



\end{document}